\documentclass{article}
\usepackage{spconf,amsmath,epsfig}
\usepackage{amsf onts,amssymb}
\usepackage{xcolor}
\usepackage{cite}
\usepackage{stfloats}
\usepackage{stmaryrd}
\usepackage{float}
\usepackage{textcomp}
\usepackage{multirow}
\usepackage{bm}
\usepackage[english]{babel}

\usepackage{hyperref}
\usepackage{subcaption}
\newcommand{\bqn}{\begin{eqnarray}}
	\newcommand{\eqn}{\end{eqnarray}}
\newcommand{\bq}{\begin{eqnarray*}}
	\newcommand{\eq}{\end{eqnarray*}}

\newcommand{\ba}{\left( \begin{array}}
\newcommand{\ea}{\end{array} \right)}

\usepackage{hyperref}
\hypersetup{
    colorlinks=true,%
    citecolor=black,%
    filecolor=blue,%
    linkcolor=red,%
    urlcolor=black,}

\usepackage[normalem]{ulem}

\title{Hodge-Decomposition  of Brain Networks}
\name{
D Vijay Anand$^1$, Moo K. Chung$^2$\thanks{Published in the \emph{2024 IEEE International Symposium on Biomedical Imaging (ISBI)}, pp.~1--5.
This arXiv version corrects spelling, grammatical, and notational errors.
This study is funded by NIH EB028753, MH133614 and NSF DMS-2010778. 
}}

\address{$^1$ University College London, London, UK\\
		$^2$ University of Wisconsin, Madison, USA\\
		{\tt ucahvad@ucl.ac.uk,  mkchung@wisc.edu}
		}

\begin{document}
\maketitle

\begin{abstract}
We analyze brain networks by decomposing them into three orthogonal components: gradient, curl, and harmonic flows, through the Hodge decomposition, a technique advantageous for capturing complex topological features. A Wasserstein distance based topological inference is developed to determine the statistical significance of each component. The Hodge decomposition is applied to human brain networks obtained from a resting-state fMRI study. Our results indicate statistically significant differences in the topological features between male and female brain networks. 
\end{abstract}

\section{Introduction}

Traditional graph models employed in functional magnetic resonance imaging (fMRI) studies have predominantly focused on capturing pairwise interactions between brain regions \cite{lee.2012.TMI,sporns.2003}. This focus often results in the neglect of higher-order interactions, which are crucial for a comprehensive understanding of brain network topology \cite{battiston.2020,sizemore2019importance}. While the significance of higher-order interactions has been increasingly acknowledged, research in this direction has been constrained by challenges such as computational complexity and the lack of effective analytical tools. To overcome these challenges, topological data analysis (TDA) has emerged as a promising technique. Persistent homology (PH), a specialized technique within TDA, leverages simplicial complexes to encode higher-order interactions in a systematic and computationally efficient manner \cite{giusti2016two,lee2011computing}. Simplicial complexes consist of simplices, each corresponding to a different level of interaction within the network. Specifically, nodes represent 0-way interactions, edges correspond to 1-way interactions, triangles denote 2-way interactions, and tetrahedra signify 3-way interactions. PH allows for a nuanced representation of networks across different spatial resolutions, thereby enriching brain imaging data \cite{chung.2019.NN,sizemore2019importance}.

PH quantifies multiscale topological features of data  through a filtration process \cite{chung.2017.CNI}. Hodge theory provides a unified framework combining simplicial homology and spectral geometry, offering insights into network topology \cite{lim.2020, barbarossa2020topological, anand.2023.TMI}. While the Hodge Laplacian, a generalization of the graph Laplacian, offers insights into the topological features of higher order simplices, the Hodge decomposition allows us to establish relationships between simplices of different dimensions \cite{barbarossa2020topological}. Hodge decomposition breaks data defined on edges (edge flow) into three orthogonal components: gradient, curl, and harmonic flows, each providing unique topological insights. The gradient flow, driven by node gradients, represents the network's gradient-like behavior. The curl flow, arising from triangle-induced flows, captures rotational patterns, while the harmonic flow exposes loop structures and topological signatures \cite{barbarossa2020topological}. Using a Wasserstein distance-based statistical approach on each component, this study assesses the topological similarities and differences between loop and non-loop flows. Further, leveraging on the properties of the decomposed networks, the study seeks to elucidate the most discriminating topological disparities in female and male  functional brain networks.

\section{Method}

\subsection{Boundary and coboundary operators}

A simplicial complex is a collection of simplices that includes nodes ($0$-simplices), edges ($1$-simplices), triangles ($2$-simplices) and their higher-dimensional counterparts. The $0$-skeleton of a simplicial complex consists only of nodes, while a $1$-skeleton comprises both nodes and edges. Graphs are examples of $1$-skeletons. A $k$-chain is a formal linear combination of $k$-simplices. The set of all such $k$-chains constitutes a group, denoted as $\mathcal{K}_k$. A sequence of these groups forms  a chain complex. To relate different chain groups, boundary operators are used.

For two successive chain groups, $\mathcal{K}_k$ and $\mathcal{K}_{k-1}$, the boundary operator $\partial_k: \mathcal{K}_k \longrightarrow \mathcal{K}_{k-1}$ for a given $k$-simplex $\sigma_k$ is defined as
\begin{equation}
\partial_k(\sigma_k) = \sum_{i=0}^k(-1)^i(v_0, \cdots, \widehat{v_i}, \cdots, v_k),
\label{eqn:boundary_map}
\end{equation}
where $(v_0, \cdots, \widehat{v_i}, \cdots, v_k)$ represents the $(k-1)$-faces of $\sigma_k$, obtained by omitting the vertex $\widehat{v_i}$. The boundary operator's matrix representation, $\mathbb{B}_k = (\mathbb{B}_k^{ij})$, is given by
\begin{equation}
\mathbb{B}_k^{ij} =
\begin{cases}
1, & \text{if } \sigma^i_{k-1} \subset \sigma^j_{k} \text{ and } \sigma^i_{k-1} \sim \sigma^j_{k}, \\
-1, & \text{if } \sigma^i_{k-1} \subset \sigma^j_{k} \text{ and } \sigma^i_{k-1} \nsim \sigma^j_{k}, \\
0, & \text{otherwise},
\end{cases}
\label{eq:boundarymatrix}
\end{equation}
where $\sim$ and $\nsim$ indicate similar and dissimilar orientations, respectively \cite{edelsbrunner.2010,huang.2023}.

Coboundary operators $\delta_k$ are duals of the boundary operators, mapping \( k \)-cochains to \( (k+1) \)-cochains:  \(\delta_k: \mathcal{K}^{k} \to \mathcal{K}^{k+1}\). The coboundary operator is the adjoint of the boundary operator \(\partial_{k+1}\). The matrix representation of \(\delta_k\) is the transpose of the matrix representation of \(\partial_{k+1}\):
$
\delta_k = \mathbb{B}_{k+1}^\top.
$

\subsection{Hodge Decomposition}

Let  \( \mathcal{C}^k \) be the space of functions  over $k$-simplices. The Hodge decomposition separates an edge flow $X \in \mathcal{C}^1$ into three orthogonal components: gradient $X_G$, curl $X_C$ and harmonic flows $X_H$:
\begin{equation}
X = X_G + X_C + X_H = \delta_0 s + \delta_1^T \bm{\phi} + X_H
\label{Eq:HHD_HL_B}
\end{equation}
with potential functions \( s  \in \mathcal{C}^0 \) and \( \bm{\phi} \in \mathcal{C}^2 \). 
The components are determined through orthogonal projections onto their respective subspaces.
\( X_G \) represents the function on the 1-chain  that can be described as the gradient of a scalar field. Similarly, \( X_C \) captures the component that behaves like the curl of a vector field, and \( X_H \) denotes the harmonic component, which is divergence-free and curl-free. $X_G$ and $X_C$ are obtained by minimizing the residual in the projection as
$$
X_G = \delta_0 \arg\min_{s \in \mathcal{C}^0} \|X-\delta_0 s\|,  X_C = \delta_1^T \arg\min_{\bm{\phi} \in \mathcal{C}^2} \|X-\delta_1^T\bm{\phi}\|.
$$
The harmonic component $X_H = X - (X_G + X_C) $ is obtained as the residual. The sum of the curl and harmonic components forms the loop flow while the gradient component is referred to as the non-loop flow.  

\begin{figure}[t!]
	\centering
	\includegraphics[width=1.0\linewidth]{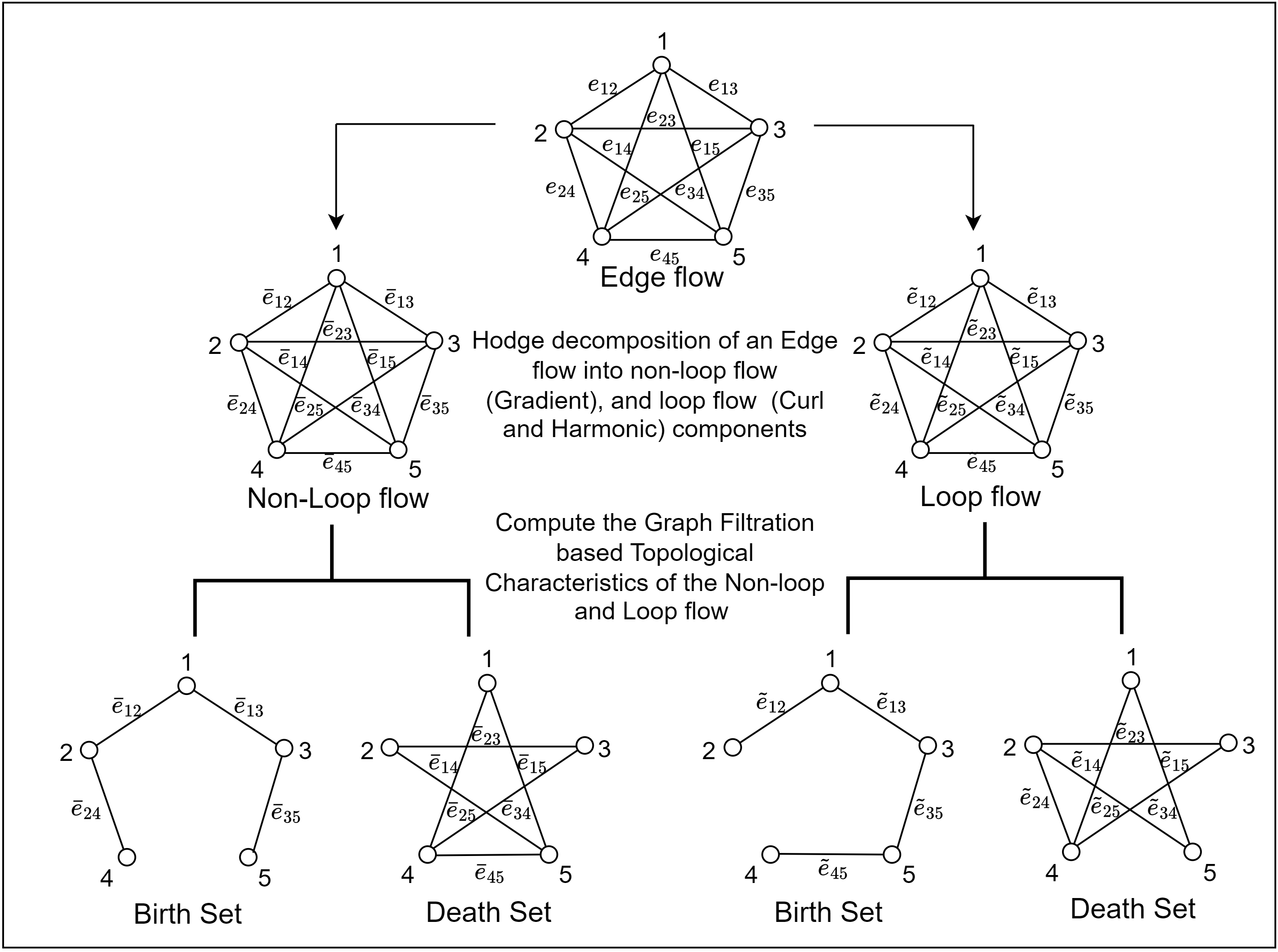}
	\caption{Illustration of the Hodge decomposition, which decomposes the edge flow into non-loop and loop flows. These networks are then separately subjected to birth-death decomposition to obtain the topological features.}
	\label{Fig:HD_Flowchart}
\end{figure}

\begin{figure}
	\centering
	\includegraphics[width=1.0\linewidth]{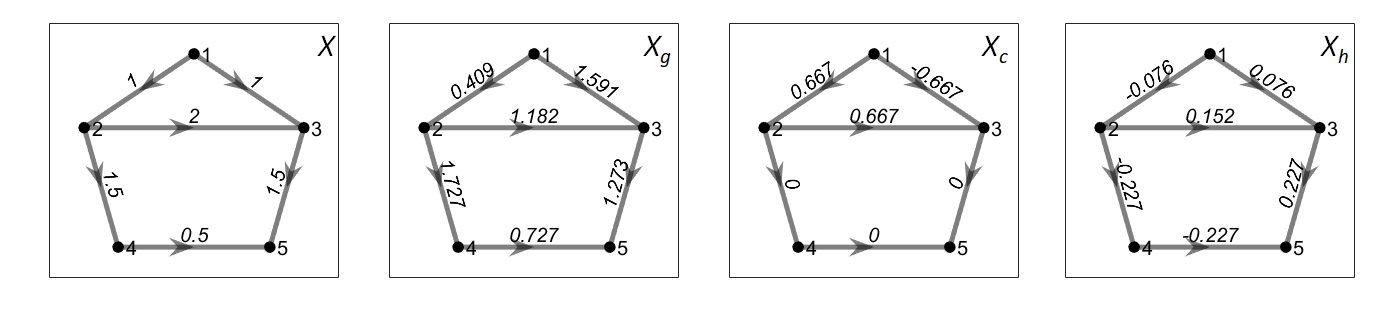}
	\caption{Hodge decomposition on a graph having 5 nodes and 6 edges. The edge flow is decomposed into gradient, curl and harmonic components.}
	\label{Fig:HD_Example}
\end{figure}

Figures \ref{Fig:HD_Flowchart} and \ref{Fig:HD_Example} illustrate the Hodge decompositions of a complete graph and a non-complete graph, respectively. The MATLAB code for performing Hodge decomposition in the least squares fashion is available at \\\url{https://github.com/laplcebeltrami/hodge}.


\subsection{Topological Inference on the Hodge Decomposition}
\label{Sec:StatinferenceWD}
To measure topological distance between graphs, we employ the {\em birth-death decomposition} (BDD), which
partitions graphs into topologically distinct subgraphs\cite{song.2021.MICCAI,song.2023}. We first apply graph filtration, a technique involving the sequential removal of edges from a graph $G$, starting with the smallest edge weight and progressing to the largest \cite{lee2011computing,chung.2017.CNI}. We identify the birth set \( B(G) \), associated with the emergence of connected components, by computing the maximum spanning tree (MST) of \( G \) using Kruskal’s or Prim’s algorithms \cite{lee2011computing}. The death set \( D(G) \) then consists of the edges not present in \( B(G) \) (Figure \ref{Fig:HD_Flowchart}), which consists of death values of cycles (loops) during the filtration. We perform BDD  independently on both non-loop and loop flows, allowing us to characterize the topology of each component of the Hodge decomposition. To measure the topological disparities between components, we use the Wasserstein distance applied to their respective BDD. Wasserstein distance provides optimal matchings that are stable to infinitesimal noise and provide robustness \cite{cohensteiner.2007, song.2023}. 

We evaluate the difference between two groups of networks \( \Omega = \{\Omega_1, \Omega_2, \ldots , \Omega_m\} \) and \( \Psi = \{\Psi_1, \Psi_2, \ldots , \Psi_n\} \). The Wasserstein distance is used as a test statistic \cite{su.2015,shi.2019,anand.2023.TMI}:
\begin{align}
\mathfrak{L}_\infty(\Omega, \Psi) &= \mathfrak{L}_{\infty}^b(\Omega, \Psi) + \mathfrak{L}_{\infty}^d(\Omega, \Psi) \nonumber \\
&= \max_{1 \leq j \leq q_0} | \bar{b}^{\Omega}_j - \bar{b}^{\Psi}_{j} | + \max_{1 \leq j \leq q_1} | \bar{d}^{\Omega}_j - \bar{d}^{\Psi}_{j} |.
\label{Eq:maxgapteststat}
\end{align}
\( \mathfrak{L}_{\infty}^b(\Omega, \Psi) \) computes the \( \infty \)-Wasserstein distance, where  $\bar{b}^{\Omega}_j$ and $\bar{b}^{\Psi}_j$ are the means of the $j$-th smallest birth values of  connected components in $\Omega$ and  $\Psi$. \( \mathfrak{L}_{\infty}^d(\Omega, \Psi) \) computes the \( \infty \)-Wasserstein distance, where 
$\bar{d}^{\Omega}_j$ and $\bar{d}^{\Psi}_j$ are the means of the $j$-th smallest death values of cycles in $\Omega$ and  $\Psi$ \cite{das.2023}. $q_0$ and $q_1$ are the number of birth and death values respectively. Under the null hypothesis of topological equivalence between the two groups, we expect \( \mathfrak{L}_\infty(\Omega, \Psi) \) to be close to zero. Deviations from this value would suggest a topological discrepancy between \( \Omega \) and \( \Psi \). Given that the null distribution of \( \mathfrak{L}_\infty(\Omega, \Psi) \) is not pre-established, we propose to approximate it through a permutation test, from which we subsequently derive the \( p \)-value.

\section{Validations}
\begin{figure}[t!]
	\centering
	\includegraphics[width=1\linewidth]{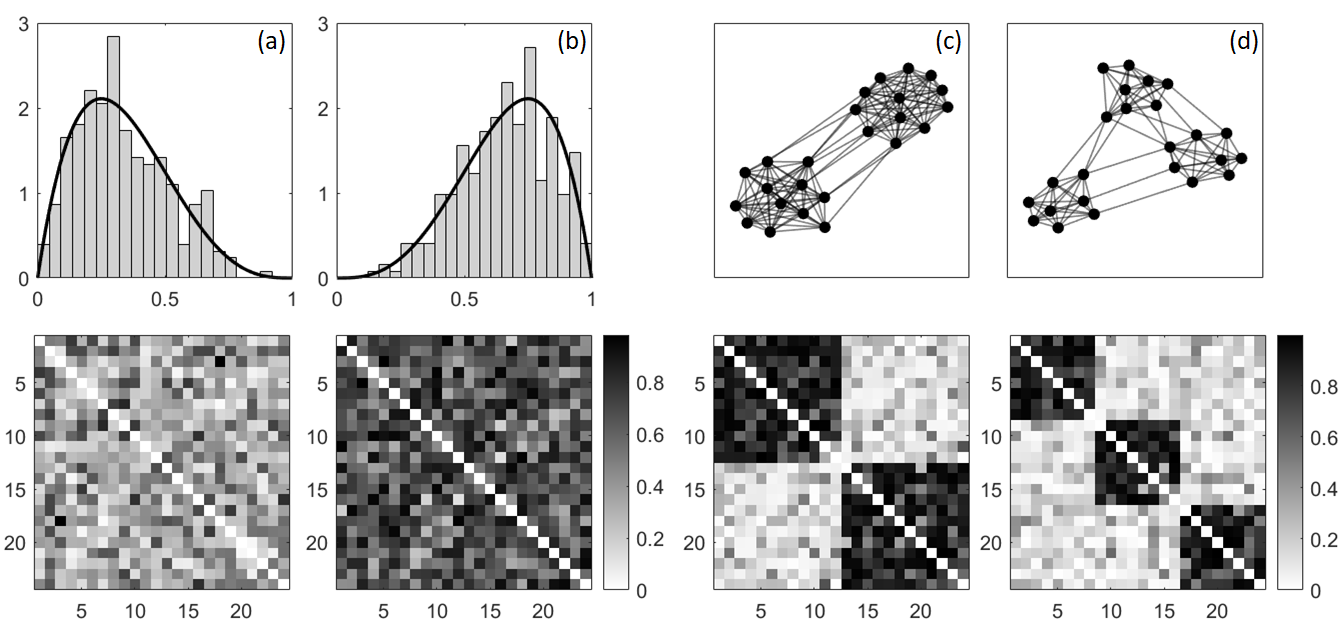}
	\caption{Edge weights following the Beta distributions with parameters (a) $(\alpha = 2, \beta = 4)$ (b) $(\alpha = 4, \beta = 2)$ with their corresponding connectivity matrices. The modular graphs obtained using the Beta distributions with (c) two and (d) three modules. The networks are thresholded at 0.4 to enable better display of the modules.}
	\label{Fig:HD_rndnet_betadist}
\end{figure}

Our method is validated in random simulations with the ground truth. We generated random modular networks with edge weights drawn from a Beta distribution \cite{das.2023}. The Beta distribution $Beta(\alpha, \beta)$ is defined on the interval $[0, 1]$ and is parameterized by positive shape parameters $\alpha$ and $\beta$. These parameters allow us to create networks with varying strengths of connectivity for comparison of their topological similarity and dissimilarity. We used  $(\alpha, \beta) = (2, 4), (4, 2)$ to generate random networks (Figure \ref{Fig:HD_rndnet_betadist}). To construct a modular network, we assigned edge weights within the same module from $Beta(\alpha, \beta)$ where $\alpha > \beta$, and edge weights between different modules from $Beta(\beta, \alpha)$. We set the number of nodes $p= 12, 18, 24$, and the number of modules $c = 2, 3, 6$, ensuring an even distribution of nodes among the modules. We employed the $\infty$-Wasserstein distance proposed in Section \ref{Sec:StatinferenceWD} to evaluate the topological similarity and dissimilarity in a 
two-group comparison setting, where each group consists of 10 networks. Statistical inferences were done using a permutation test with 100,000 permutations. We repeated the simulations independently 10 times, reporting the average \( p \)-value in Table \ref{Table:SimulationStudyRMN_Topodist}, where top rows test false negatives while the bottom rows test for false positives.

\begin{table}[t]
\caption{The performance results of the Wasserstein distance on the edge flows, loop and non-loop components. Smaller \( p \)-values are better when there are network differences (top rows) and larger \( p \)-values are better when there are  no network differences (bottom rows).}
\label{Table:SimulationStudyRMN_Topodist}
\centering
\setlength{\tabcolsep}{1pt}
\begin{tabular}{c| c | c c c}
	Nodes & Modules &   \multicolumn{3}{c}{Topological Inference} \\
	\hline
	$p$   & $c$ & Edge flow ~~~   &  Loop flow~~~  &   Non-loop flow\\
	\hline
	12 vs. 12 & 2 vs. 3 & 0.0001  & 0.0039 & 0.0000\\
			  & 3 vs. 6 & 0.0011  & 0.0005 & 0.0002\\
	18 vs. 18 & 2 vs. 3 & 0.0000  & 0.0012 & 0.0000\\
			  & 3 vs. 6 & 0.0003  & 0.0001 & 0.0001\\
	24 vs. 24 & 2 vs. 3 & 0.0000  & 0.0001 & 0.0000\\
			  & 3 vs. 6 & 0.0000  & 0.0001 & 0.0000\\
	\hline
	24 vs. 24 & 2 vs. 2 & 0.1135  & 0.9669 & 0.1794\\
			  & 3 vs. 3 & 0.5348  & 0.7451 & 0.8864 \\
			  & 6 vs. 6 & 0.2863  & 0.4055 & 0.6055 \\
	\hline
\end{tabular}
\end{table}

The proposed $\infty$-Wasserstein distance-based test statistic exhibits robust performance on both the loop and non-loop flows. The $\mathfrak{L}_{\infty}$ distance effectively discriminated networks in both the non-loop (gradient) and loop (curl) components when network differences were present (as shown in the top rows of the table). In scenarios with no network differences (bottom rows of the table), both the loop and non-loop flows yielded satisfactory results. This underscores that the modularity in the network is aptly captured by both the non-loop and loop components of the Hodge decomposition, and that our $\infty$-Wasserstein distance is capable of discerning variations in modularity.

\section{Application}
\subsection{Functional brain imaging data and preprocessing}
\begingroup
\setlength{\emergencystretch}{1em}
We used the resting-state fMRI (rs-fMRI) in the Human Connectome Project \cite{vanessen.2012}. rs-fMRI data are collected at 2 mm isotropic voxels and 1200 time points. Data that was subjected to the standard minimal preprocessing pipelines \cite{glasser.2016} was used. Volumes with framewise displacement larger than $0.5$mm and their neighbors were scrubbed \cite{vanessen.2012}. Excessive head movement was excluded from the study. Subsequently, the Automated Anatomical Labeling (AAL) template is used to parcellate and average rs-fMRI  spatially into 116 non-overlapping anatomical regions. The details on image processing are given in \cite{huang.2020.NM}. The final data is comprised of the fMRI of 400 subjects of which 168 are males and 232 are females. Subsequently, Pearson correlation is used in computing a 116 $\times$ 116 correlation matrix per subject. The Hodge decomposition considers both positive and negative edge values with appropriate direction of orientation. Thus, we did not discard negative correlations.  
\par
\endgroup

\begin{figure*}[t] 
	\centering
	\includegraphics[width=\linewidth]{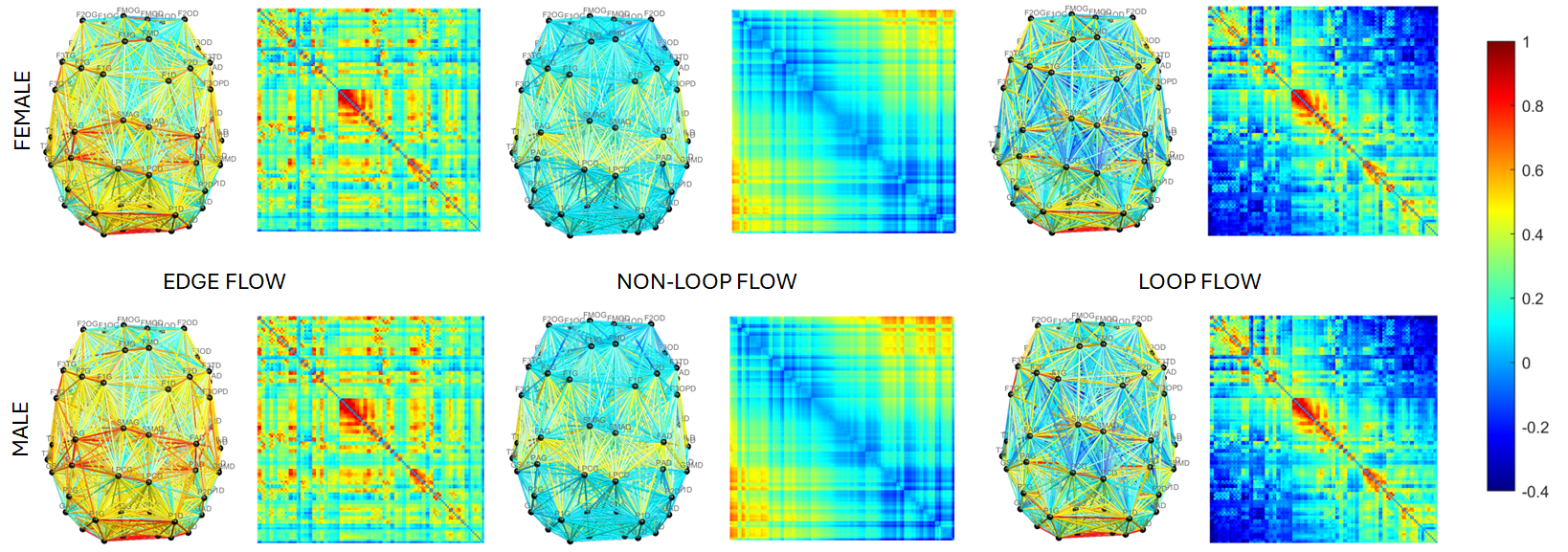}
	\caption{The average connectivity (edge flow), non-loop flow (middle), which is gradient flow $X_G$, and the loop flow (right), which is the sum of curl  $X_C$  and harmonic flow $X_H$, of the female (top) and male networks (bottom) along with their connectivity matrices. The colorbar represents edge weight values for each flow type.}
	\label{Fig:Brain_X_FemaleMale}
\end{figure*}

\subsection{Hodge decomposition of the brain network}

We decomposed individual brain networks using the Hodge decomposition. In Figure \ref{Fig:Brain_X_FemaleMale}, the Hodge decomposition applied to average female and male brain networks is displayed. We then assessed if there are topological differences between females and males in the original connectivity (edge flow). Following the test procedure in Section \ref{Sec:StatinferenceWD}, the Wasserstein distances $\mathfrak{L}_{\infty}^b$ on birth values for testing 0D topology difference and $\mathfrak{L}_{\infty}^d$ on death values for testing 1D topology difference are separately used. The permutation test conducted on both the birth set (first term) and the death set (second term) yielded 
$p$-values of 0.0177 and 0.0110, respectively. These results indicate the presence of significant topological differences in the original connectivity matrices.

We further determined if we can detect topological differences in the decomposed components (Figure \ref{Fig:HD_Example}). The gradient component sums to zero along any cycle. The curl components are zero for edges that are not a 2-simplex boundary and the entries sum to zero around each node. The harmonic component sums to zero around each node, and it also sums to zero along each 2-simplex. We tested the topological equivalence of female brain networks and male brain networks using the Wasserstein distance  (\ref{Eq:maxgapteststat}). 
We first considered the  network constructed from the gradient component and performed the birth death decomposition. We then carried out the permutation test (\( p \)-value $=$ 0.008). The gradient flow reflects a directed transmission of information or activity across the brain, indicative of a potential difference between nodes that drives the flow from areas of high to low functional connectivity strength. Also we performed the permutation test on the curl component (\( p \)-value $=$ 0.0296). Curl flows are akin to vortices, where the functional connectivity follows circular patterns, suggesting a rotational or cyclic exchange of information between brain regions, a phenomenon that plays a pivotal role in the brain's ability to process and integrate information across various cognitive domains.

We also tested if we can detect 0D and 1D topological signals separately
in each component. For the non-loop component (gradient flow), the permutation test yielded $p$-values of 0.0088 and 0.0080 for the birth (0D topology) and death (1D topology) values, respectively. The birth values, representing the emergence of new connections, and the death values, indicating the dissolution of existing connections, both demonstrate the dynamic nature of the brain's connectivity that does not necessarily form closed loops. For loop components, the permutation test yielded $p$-values of 0.0019 for the birth values and 0.1582 for the death values, indicating a highly significant presence of loop flows for birth values but not for death values. Loop flows, which include curl and harmonic flows, characterize the cyclic and recurrent connectivity patterns. The significant result for the birth values suggests that the formation of cyclic patterns is a notable feature of brain connectivity, supporting the brain's integrative and cooperative processes. However, the non-significant result for the death values hints at the stability of these looped connections once established, reflecting the brain's tendency to maintain certain cyclic patterns over time.

\section{Conclusion}
This study detailed the use of Hodge decomposition—gradient, curl, and harmonic components—in topological data analysis of brain networks. Applying this technique to rs-fMRI correlation networks, we subsequently categorized these components into loop and non-loop flows. By incorporating the Wasserstein distance, we differentiated network types, showcasing the components' ability to detect topological differences. Tested on static functional brain networks, our method discriminated between male and female brain networks. The extension to dynamically changing brain networks and time-dependent topological changes is left as future work.

\bibliographystyle{IEEEbib}
\bibliography{reference.ISBI2024,reference.2023.10.03moo} 

\begin{thebibliography}{10}

\bibitem{lee.2012.TMI}
H.~Lee, H.~Kang, M.K. Chung, B.-N. Kim, and D.S Lee,
\newblock ``Persistent brain network homology from the perspective of
  dendrogram,''
\newblock {\em IEEE Transactions on Medical Imaging}, vol. 31, pp. 2267--2277,
  2012.

\bibitem{sporns.2003}
O.~Sporns,
\newblock {\em Graph Theory Methods for the Analysis of Neural Connectivity
  Patterns}, pp. 171--185,
\newblock Springer US, Boston, MA, 2003.

\bibitem{battiston.2020}
F.~Battiston, G.~Cencetti, I.~Iacopini, V.~Latora, M.~Lucas, A.~Patania, J.-G.
  Young, and G.~Petri,
\newblock ``Networks beyond pairwise interactions: structure and dynamics,''
\newblock {\em Physics Reports}, vol. 874, pp. 1--92, 2020.

\bibitem{sizemore2019importance}
A.E. Sizemore, J.E. Phillips, R.~Ghrist, and D.~S. Bassett,
\newblock ``The importance of the whole: topological data analysis for the
  network neuroscientist,''
\newblock {\em Network Neuroscience}, vol. 3, no. 3, pp. 656--673, 2019.

\bibitem{giusti2016two}
C.~Giusti, R.~Ghrist, and D.S. Bassett,
\newblock ``Two’s company, three (or more) is a simplex,''
\newblock {\em Journal of Computational Neuroscience}, vol. 41, no. 1, pp.
  1--14, 2016.

\bibitem{lee2011computing}
H.~Lee, M.K. Chung, H.~Kang, B.~Kim, and D.S. Lee,
\newblock ``Computing the shape of brain networks using graph filtration and
  {G}romov-{H}ausdorff metric,''
\newblock in {\em International Conference on Medical Image Computing and
  Computer-Assisted Intervention}. Springer, 2011, pp. 302--309.

\bibitem{chung.2019.NN}
M.K. Chung, H.~Lee, A.~DiChristofano, H.~Ombao, and V.~Solo,
\newblock ``Exact topological inference of the resting-state brain networks in
  twins,''
\newblock {\em Network Neuroscience}, vol. 3, pp. 674--694, 2019.

\bibitem{chung.2017.CNI}
M.K. Chung, H.~Lee, V.~Solo, R.J. Davidson, and S.D. Pollak,
\newblock ``Topological distances between brain networks,''
\newblock {\em International Workshop on Connectomics in Neuroimaging}, vol.
  10511, pp. 161--170, 2017.

\bibitem{lim.2020}
L.H. Lim,
\newblock ``{H}odge {L}aplacians on graphs,''
\newblock {\em SIAM Review}, vol. 62, no. 3, pp. 685--715, 2020.

\bibitem{barbarossa2020topological}
S.~Barbarossa and S.~Sardellitti,
\newblock ``Topological signal processing over simplicial complexes,''
\newblock {\em IEEE Transactions on Signal Processing}, vol. 68, pp.
  2992--3007, 2020.

\bibitem{anand.2023.TMI}
D.V. Anand and M.K. Chung,
\newblock ``{Hodge-Laplacian} of brain networks,''
\newblock {\em IEEE Transactions on Medical Imaging}, vol. 42, pp. 1563--1473,
  2023.

\bibitem{edelsbrunner.2010}
H.~Edelsbrunner and J.~Harer,
\newblock {\em Computational topology: {A}n introduction},
\newblock American Mathematical Society, 2010.

\bibitem{huang.2023}
J.~Huang, M.K. Chung, and A.~Qiu,
\newblock ``Heterogeneous graph convolutional neural network via
  hodge-laplacian for brain functional data,''
\newblock in {\em International Conference on Information Processing in Medical
  Imaging}. Springer, 2023, pp. 278--290.

\bibitem{song.2021.MICCAI}
T.~Songdechakraiwut, L.~Shen, and M.K. Chung,
\newblock ``Topological learning and its application to multimodal brain
  network integration,''
\newblock {\em Medical Image Computing and Computer Assisted Intervention
  (MICCAI)}, vol. 12902, pp. 166--176, 2021.

\bibitem{song.2023}
T.~Songdechakraiwut and M.K Chung,
\newblock ``Topological learning for brain networks,''
\newblock {\em Annals of Applied Statistics}, vol. 17, pp. 403--433, 2023.

\bibitem{cohensteiner.2007}
D.~Cohen-Steiner, H.~Edelsbrunner, and J.~Harer,
\newblock ``Stability of persistence diagrams,''
\newblock {\em Discrete and Computational Geometry}, vol. 37, pp. 103--120,
  2007.

\bibitem{su.2015}
Z.~Su, W.~Zeng, Y.~Wang, Z.-L. Lu, and X.~Gu,
\newblock ``Shape classification using wasserstein distance for brain
  morphometry analysis,''
\newblock in {\em International Conference on Information Processing in Medical
  Imaging}. Springer, 2015, pp. 411--423.

\bibitem{shi.2019}
J.~Shi and Y.~Wang,
\newblock ``Hyperbolic wasserstein distance for shape indexing,''
\newblock {\em IEEE Transactions on Pattern Analysis and Machine Intelligence},
  vol. 42, pp. 1362--1376, 2019.

\bibitem{das.2023}
S.~Das, D.V. Anand, and M.K. Chung,
\newblock ``Topological data analysis of human brain networks through order
  statistics,''
\newblock {\em PLOS One}, vol. 18, no. 3, pp. e0276419.

\bibitem{vanessen.2012}
D.C. Van~Essen, K.~Ugurbil, E.~Auerbach, D.~Barch, T.E.J. Behrens, R.~Bucholz,
  A.~Chang, L.~Chen, M.~Corbetta, and S.W. Curtiss,
\newblock ``The human connectome project: a data acquisition perspective,''
\newblock {\em NeuroImage}, vol. 62, pp. 2222--2231, 2012.

\bibitem{glasser.2016}
M.F. Glasser, S.M. Smith, D.S. Marcus, J.L.R. Andersson, E.J. Auerbach, T.E.J.
  Behrens, T.S. Coalson, M.P. Harms, M.~Jenkinson, and S.~Moeller,
\newblock ``The human connectome project's neuroimaging approach,''
\newblock {\em Nature Neuroscience}, vol. 19, pp. 1175, 2016.

\bibitem{huang.2020.NM}
S.-G. Huang, S.-T. Samdin, C.M. Ting, H.~Ombao, and M.K. Chung,
\newblock ``Statistical model for dynamically-changing correlation matrices
  with application to brain connectivity,''
\newblock {\em Journal of Neuroscience Methods}, vol. 331, pp. 108480, 2020.

\end{thebibliography}
\end{document}